\documentclass[aps,pre,reprint,10pt,twocolumn,showpacs,nofootinbib]{revtex4-1} 

\usepackage{amsmath}
\usepackage{amsfonts}
\usepackage{amssymb}
\usepackage{graphicx}
\usepackage{subfigure}
\usepackage[pdfborder={0 0 0}]{hyperref}  
\usepackage{color}
\usepackage[latin1]{inputenc}
\usepackage{float}
\usepackage{comment}

\begin{document}

\title{Thermal entanglement in $2\otimes3$ Heisenberg chains via distance between states}
 
\begin{abstract}
Most of the work involving entanglement measurement focuses on systems that can be modeled by two interacting qubits. This is due to the fact that there are few studies presenting entanglement analytical calculations in systems with spins $s > 1/2$. In this paper we present for the first time an analytical way of calculating thermal entanglement in a dimension $2\otimes3$ Heisenberg chain through the distance between states. We use the Hilbert-Schmidt norm to obtain entanglement. The result obtained can be used to calculate entanglement in chains with spin-$1/2$ coupling with spin-$1$, such as ferrimagnetic compounds as well as compounds with dimer-trimer coupling. 
\end{abstract}

\author{Saulo Luis Lima da Silva}

\email{saulosilva@cefetmg.br}

\affiliation{Centro Federal de Educa\c c\~ao Tecnol\'ogica de Minas Gerais,\\
Avenida monsenhor Luiz de Gonzaga, 103 -- centro -- 37250-000 --
Nepomuceno -- MG -- Brazil.}

\maketitle

\section{Introduction}
Since the end of the last century we have seen with great enthusiasm the great achievements in the field of quantum information. Remarkably, much effort was devoted to the understanding, measurement, and control of entanglement \cite{ref1, ref2, ref3, ref4, ref5, ref6, ref7, Hauke,ref8, Mathew}. Despite this, we have not yet been able to obtain a unique form of characterization and measure of entanglement. What we have are criteria that, in some cases, allow us to determine if there is entanglement in the system. A major breakthrough was struck by the work of the Horodecki family in 1996 by showing what became known as the Pers-Horodecki criterion \cite{ref2}. They showed that for systems whose dimension in the Hilbert space is not greater than $2\otimes3$, the positivity of the partial transpose of the reduced density matrix of the system is necessary and sufficient condition for the entanglement.
   
The Peres-Horodecki criterion allows us to qualitatively determine entanglement. There are several proposals in the literature for the quantitative determination of entanglement \cite{ref3, ref4, ref5, ref9, ref10}. One of the best known is the formation entanglement, which in the case of two qubits lies in the well-known Wootters formula \cite{ref5}. Although there are generalizations of concurrence for $s>1/2$\cite{You, Osterloh, Bahmani}, there are so far no simple expressions \cite{Scheie}. A more general proposal is the distance between states as a measure of entanglement \cite{ref3}. In this case, the distance between the state of interest and the set of separable states is used as a measure of the degree of entanglement of the system. This proposal has a strong geometric appeal and the advantage of not being limited to the size of the system or the number of particles of the system. The disadvantage is that numerical methods are usually required for the calculation of the distance between states, since the determination of the separable matrix set element closest to the state of interest is not trivial. 

To quantify entanglement in $2\otimes3$ systems, negativity is generally used \cite{ref13,ref14}. However, as demonstrated by Eisert et al. \cite{ref15}, it is not a good measure since it does not induce the same ordering as the entanglement of formation and cannot, in general, be used to determine the most entangled state for a given family of density operators.

In recent work, we have shown an analytical way to quantitatively calculate thermal and macroscopic entanglement using the distance between states \cite{ref10}. For this, we made use of the Peres-Horodecki criterion, which allows us to analyze systems larger than $2\otimes2$. We use the Hilbert-Schmidt norm as a measure of the distance between states. Later, other works were published using this technique to measure spin chain entanglement of various compounds with and without the application of magnetic field \cite{ref10,ref11,ref12}.
   
In this paper we aim to present for the first time an analytical way of calculating thermal entanglement in a dimension $2\otimes3$ Heisenberg chain through the distance between states. We use the Hilbert-Schmidt norm to obtain entanglement. The result obtained can be used to calculate entanglement in chains with spin-$1/2$ coupling with spin-$1$, such as ferrimagnetic compounds as well as compounds with dimer-trimer coupling.

The outline of this letter is as follows. In Section II we describe the physical model. In Section III, the partition function and the system density operator are calculated. The analytical calculation of the entanglement of the system is done in Section IV. Section V is dedicated to the conclusions.

\section{The model}

The model consists of a one-dimensional chain of spin-$1/2$ and spin-$1$ alternating with ferrimagnetic coupling between first neighbors.
This system can be modeled by the following Hamiltonian
\begin{equation}
\mathcal{H}=-J\sum_{i}\vec{s}_{i} \cdot \vec{S}_{i+1},
\end{equation}
where $J$ is the exchange coupling constant, $s_i$ is the spin-$1/2$ operator on site $i$ and $S_{i+1}$ is the spin-$1$ operator on site $i+1$.
In matrix notation we can write $s_i$ in terms of the Pauli matrices, $\vec{s}=\frac{1}{2}\vec{\sigma}=\frac{1}{2}(\sigma^x,\sigma^y,\sigma^z)$ 
and $S$ in the form $\vec{S}=\vec{I}=(I^x,I^y,I^z)$ where
\[
I^x= \frac{1}{\sqrt{2}} \left[\begin{array}{ccc} 
0 & 1 & 0 \\
1 & 0 & 1 \\
0 & 1 & 0
\end{array}\right],
\]
\[
I^y= \frac{i}{\sqrt{2}} \left[\begin{array}{ccc} 
0 & -1 & 0 \\
1 & 0 & -1 \\
0 & 1 & 0
\end{array}\right],
\]
\[
I^z=\left[\begin{array}{ccc} 
1 & 1 & 0 \\
1 & 0 & 0 \\
0 & 0 & -1
\end{array}\right].
\]
Thus
\[
\mathcal{H}= -\frac{J}{2}\left[\begin{array}{cccccc} 
1 & 0 & 0 & 0 & 0 & 0 \\
0 & 0 & 0 & \sqrt{2} & 0 & 0 \\
0 & 0 & -1 & 0 & \sqrt{2} & 0 \\
0 & \sqrt{2} & 0 & -1 & 0 & 0 \\
0 & 0 & \sqrt{2} & 0 & 0 & 0 \\
0 & 0 & 0 & 0 & 0 & 1
\end{array}\right].
\]
The eigenvalues of hamiltonian are 
$$\lambda_{1,2,5,6}= -\frac{J}{2}, \,\,\, \mathrm{and} \,\,\, \lambda_{3,4}= J,$$
with $J < 0.$ Taking 
$$\vert \alpha \rangle =
\begin{pmatrix}
1 \\
0
\end{pmatrix}, \,\,\, \vert \beta \rangle =
\begin{pmatrix}
0 \\
1
\end{pmatrix},$$
$$\vert 1 \rangle =
\begin{pmatrix}
	1 \\
	0 \\
	0
\end{pmatrix}, \,\,\, \vert 0 \rangle =
\begin{pmatrix}
0 \\
1 \\
0
\end{pmatrix}, \,\,\, \vert -1 \rangle=
\begin{pmatrix}
0 \\
0 \\
1
\end{pmatrix},$$
the respective eigenvectors are
\[
|\phi_1 \rangle = | \beta -1\rangle, \,\,\,\,\, 
|\phi_2 \rangle = | \alpha 1\rangle,
\]
$$|\phi_3 \rangle = \frac{1}{\sqrt{3}}\left(-\sqrt{2}|\alpha -1\rangle +|\beta 0 \rangle \right),$$
$$|\phi_4 \rangle = \sqrt{\frac{2}{3}}\left(\frac{-\sqrt{2}}{2}|\alpha 0\rangle +|\beta 1 \rangle \right),$$
$$|\phi_5 \rangle = \sqrt{\frac{2}{3}}\left(\frac{\sqrt{2}}{2}|\alpha -1\rangle +|\beta 0 \rangle \right),$$
$$|\phi_6 \rangle = \frac{1}{\sqrt{3}}\left(\sqrt{2}|\alpha 0\rangle +|\beta 1 \rangle \right).$$

\section{Partition Function and the Density Operator}
In a finite temperature and thermal equilibrium system the density operator is given by $\rho=Z^{-1}e^{-\beta \mathcal{H}},$ 
where $Z=\mathrm{Tr}(e^{-\beta \mathcal{H}})$ is the partition function and $\beta =(k_BT)^{-1}.$ The density operator takes the form
\[
\rho(T)= \frac{1}{Z}\left[\begin{array}{cccccc} 
v & 0 & 0 & 0 & 0 & 0 \\
0 & x & 0 & w & 0 & 0 \\
0 & 0 & y & 0 & w & 0 \\
0 & w & 0 & y & 0 & 0 \\
0 & 0 & w & 0 & x & 0 \\
0 & 0 & 0 & 0 & 0 & v
\end{array}\right],
\]
where $$Z=2e^{\frac{J}{2k_BT}}\left(1+2e^{-\frac{3J}{k_BT}}\mathrm{cosh}\left( \frac{3J}{k_BT}\right)\right),$$
$$v=e^{\frac{J}{2k_BT}},$$ $$x=\frac{1}{3}\left( e^{-\frac{J}{k_BT}}+2e^{\frac{J}{2k_BT}}\right),$$
$$y=\frac{1}{3}\left( 2e^{-\frac{J}{k_BT}}+e^{\frac{J}{2k_BT}}\right),$$
$$w=\frac{\sqrt{2}}{3}\left(-e^{-\frac{J}{k_BT}}+e^{\frac{J}{2k_BT}}\right),$$
The eigenvalues of the partial transposition of the density operator are
\[
\begin{array}{cl}
\lambda_{1,2}= & \frac{1}{2}\left(v+x-\sqrt{(v-x)^2+4w^2}\right),\\
\lambda_{3,4}= & \frac{1}{2}\left(v+x+\sqrt{(v-x)^2+4w^2}\right),\\
\lambda_{5,6}= & y.
\end{array}
\]

\section{Entanglement}
According to the Peres-Horodecki criterion \cite{ref2}, the system will be entangled when $\lambda_{1,2}<0$
(the other eigenvalues are always positive) and this occurs for 
\begin{equation}\label{tri}
T<-\frac{3J}{2k_B \mathrm{ln}4}.
\end{equation}

Note that this expression is only valid for $J <0$ (coupling 
antiferrimagnetic), for $J>0$ there is no entanglement (all eigenvalues of the partial transpose are positive).
Note that the equation (\ref{tri}) gives us the critical entanglement temperature of the system
\begin{equation}\label{tcri}
T_{E}=\frac{3|J|}{2k_B \mathrm{ln}4},
\end{equation}
and this result is in perfect agreement with that obtained by Wang et al. in which they used negativity as a measure of 
entanglement \cite{ref13}. Eq. (\ref{tcri}) shows that the critical temperature in this case is higher than the 
critical temperature $T_{E1/2}$ for the a 2-qubit 1-D XXX Heisenberg, result in perfect agreement with \cite{ref13}. 
$T_{E1/2}$ is given by \cite{ref16} 
$$T_{E1/2}=\frac{\vert J \vert}{k_B\mathrm{ln}3},$$
therefore $T_{E1/2}<T_{E}.$

Now let's quantify the entanglement using the distance between states. We will use here the method presented by 
Del Cima et al. \cite{ref10}. From the Peres-Horodecki criterion we know that the system will be entangled when
\begin{equation}\label{eq1}
v+x < \sqrt{(v-x)^2+w^2},
\end{equation}
and that will happen when $T<\frac{3\vert J \vert}{2k_B\mathrm{ln}4}.$ This allows us to discriminate the set of 
density matrices separable (stisfies the Eq.\ref{eq1}) from the set of entangled matrices (does not stisfies the 
Eq.\ref{eq1}). Be
\[
\rho_s= \frac{1}{Z}\left[\begin{array}{cccccc} 
v_s & 0 & 0 & 0 & 0 & 0 \\
0 & x_s & 0 & w_s & 0 & 0 \\
0 & 0 & y_s & 0 & w_s & 0 \\
0 & w_s & 0 & y_s & 0 & 0 \\
0 & 0 & w_s & 0 & x_s & 0 \\
0 & 0 & 0 & 0 & 0 & v_s
\end{array}\right],
\]
the density operator that designates the separable states and 
\begin{equation}\label{eq1.1}
\rho_e= \frac{1}{Z}\left[\begin{array}{cccccc} 
v_e & 0 & 0 & 0 & 0 & 0 \\
0 & x_e & 0 & w_e & 0 & 0 \\
0 & 0 & y_e & 0 & w_e & 0 \\
0 & w_e & 0 & y_e & 0 & 0 \\
0 & 0 & w_e & 0 & x_e & 0 \\
0 & 0 & 0 & 0 & 0 & v_e
\end{array}\right],
\end{equation}
the density operator that designates entangled states. Via Hilbert-Schmidt norm, the distance between these two states 
is given by
{
	\fontsize{8}{8}\selectfont
\begin{align}\label{eq2}
&\mathcal{D}(\rho_s,\rho_e)=\sqrt{\mathrm{Tr}(\rho_s-\rho_e)^2}= \nonumber\\
&=\sqrt{2(v_s-v_e)^2+2(x_s-x_e)^2+2(y_s-y_e)^2+4(w_s-w_e)^2}.
\end{align}
}
The entanglement is given by the minimum of (\ref{eq2})
\begin{equation}\label{eq3}
\mathcal{E}(\rho)=\mathrm{min}\mathcal{D}(\rho_s,\rho_e).
\end{equation}
The minimum occurs when we take 
$T=\frac{3\vert J \vert }{2\mathrm{ln}4}$ in (\ref{eq1.1}).

\begin{figure}[ht!]
	\centering
	\includegraphics[width=6.5 cm]{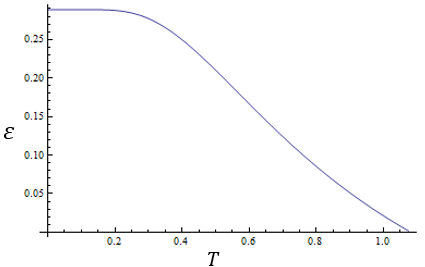}
	\caption{Entanglement as a function of temparetaura taking $J = -1$.}
	\label{f:0}
\end{figure}

In Figure \ref{f:0} we see the entanglement given by Equation (\ref{eq3}) taking $J=-1$.
We know that for the a 2-qubit 1-D XXX Heisenberg the entanglement in $T=0$ is equal to $1$.
Therefore, although the critical entanglement temperature is higher in our model, the 
entanglement is higher in 2-qubit 1-D XXX Heisenberg.

Entanglement at $T = 0$ does not depend on $J$ since  $$\lim\limits_{T \rightarrow 0}\mathcal{E}(\rho)=0.288675.$$
The entanglement at $T = 0$ can be changed by changing $S$. For $S> 1$ the entanglement at 
$T = 0$ decreases and the critical entanglement temperature increases, as shown by Wang et al. \cite{ref13}.

\section{Conclusion}

We have demonstrad a way of analytically calculating bipartite entanglement in a $2\times3$ Heisenberg chain. 
This method has the advantage of being more general than negativity and simpler than calculations involving the 
generalization of cuncurrence. We show that in such systems the entanglement at $T = 0$ is independent of the 
coupling constant and that the entanglement is more robust to changes in temperature than in $2\times2$ Heisenberg
chain. This is evident from the fact that the critical entanglement temperature is higher in the model studied 
than in the two qubits model. However, in the model studied, entanglement is generally less than in a $2\times2$ 
Heisenberg chain.

\label{sec:Conc}


\bibliographystyle{apsrev4-1}
\bibliography{ensembles}

\end{document}